\documentclass[aps,prd,twocolumn,preprintnumbers,amsmath,amssymb,nofootinbib,floatfix]{revtex4}
\usepackage[utf8]{inputenc} 
\usepackage[parfill]{parskip}    
\usepackage{graphicx}
\usepackage{amssymb}
\usepackage{epstopdf}
\usepackage{verbatim}
\usepackage{url}
\usepackage{hyperref}
\newcommand{\be}{\begin{equation}}
\newcommand{\bea}{\begin{eqnarray}}
\newcommand{\beq}[1]{\begin{equation}\label{#1}}
\newcommand{\eq}[1]{Eq.~(\ref{#1})}

\newcommand{\ee}{\end{equation}}
\newcommand{\eea}{\end{eqnarray}}
\newcommand{\eeq}{\end{equation}}
\newcommand{\Fref}[1]{Fig.~\ref{#1}}
\newcommand{\lsim}{\!\mathrel{\hbox{\rlap{\lower.55ex \hbox{$\sim$}} \kern-.34em \raise.4ex \hbox{$<$}}}}
\newcommand{\gsim}{\!\mathrel{\hbox{\rlap{\lower.55ex \hbox{$\sim$}} \kern-.34em \raise.4ex \hbox{$>$}}}}

\newcommand{\vev}[1]{\left\langle #1 \right\rangle}
\newcommand{\abs}[1]{\left| #1 \right|}
\def\eg{{\it e.g.}}
\def\ie{{\it i.e.}}
\newcommand{\sSI}{\sigma_{\rm SI}}
\newcommand{\sSD}{\sigma_{\rm SD}}
\newcommand{\GeV}{\text{ GeV}}

\begin{document}

\setlength{\baselineskip}{0.22in}

\preprint{MCTP-16-27}
\preprint{FERMILAB-PUB-16-534-T}

\title{$Z$ boson mediated dark matter beyond the effective theory}

\author{John Kearney}
\vspace{0.2cm}
\affiliation{Theoretical Physics Department,\\
Fermi National Accelerator Laboratory,\\
Batavia, IL 60510 USA}
\author{Nicholas Orlofsky and Aaron Pierce}
\vspace{0.2cm}
\affiliation{Michigan Center for Theoretical Physics (MCTP) \\
Department of Physics, University of Michigan, Ann Arbor, MI
48109}

\date{\today}

\begin{abstract}
Direct detection bounds are beginning to constrain a very simple model of weakly interacting dark matter---a Majorana fermion with a coupling to the $Z$ boson.
In a particularly straightforward gauge-invariant realization, this coupling is introduced via a higher-dimensional operator.
While attractive in its simplicity, this model generically induces a large $\rho$ parameter. An ultraviolet completion that avoids an overly large contribution to $\rho$ is the singlet-doublet model.
We revisit this model, focusing on the Higgs blind spot region of parameter space where spin-independent interactions are absent. 
This model successfully reproduces dark matter with direct detection mediated by the $Z$ boson, but whose cosmology may depend on additional couplings and states.
Future direct detection experiments should effectively probe a significant portion of this parameter space, aside from a small coannihilating region. As such, $Z$-mediated thermal dark matter as realized in the singlet-doublet model represents an interesting target for future searches.
\end{abstract}

\maketitle

\section{Introduction}

Weakly interacting massive particles (WIMPs) remain an attractive thermal dark matter (DM) candidate. 
However, while WIMPs exhibit weak scale interactions, the precise mechanism through which the DM interacts with visible matter (beyond its gravitational interactions) is unknown. One possibility is to take the ``W'' in WIMP seriously. That is, the interactions with the Standard Model (SM) are mediated not just by particles with masses near the weak scale, but by the carriers of the weak force: the $W$, $Z$ and Higgs ($h$) bosons.   It is of interest to understand the current experimental status of such models, as they represent minimal set-ups and give insight into the extent to which the WIMP paradigm is being probed.

Direct detection experiments place bounds on the spin-independent (SI) couplings of such WIMPs, which at tree level arise from exchange of the $h$ or $Z$, and their spin-dependent (SD) couplings, which at tree level arise from exchange of the $Z$. The latest bounds on SI scattering arise from  PandaX \cite{Tan:2016zwf} and LUX \cite{Akerib:2016vxi}.  
DM that interacts with the $Z$ boson via vectorial couplings, $g_{V} (\bar{\chi} \gamma_{\mu} \chi) Z^{\mu}$, is very strongly constrained; see, \eg, \cite{Essig:2007az}. For $g_{V} \sim g_Z \equiv  g_2/(2\cos{\theta_W})$, such a $\chi$ can comprise only $\lsim 10^{-6}$ of the DM.\footnote{It is possible, however, that such a tiny fraction of the DM might be observed at direct detection experiments and perhaps the LHC \cite{Halverson:2014nwa}.}  This dangerous interaction can be neatly forbidden by positing that the DM is a Majorana fermion, for which $\bar{\chi} \gamma^{\mu} \chi$ vanishes identically. This is the case, for example, for the neutralino of the Minimal Supersymmetric Standard Model (MSSM).  Majorana fermions can retain direct detection cross sections that appear at an interesting level either via SD couplings to the $Z$ boson and/or SI interactions with the Higgs boson. 

We pay special attention to regions of parameter space where the DM thermal relic abundance matches the value measured by the Planck experiment $\Omega_{DM} h^2 = 0.1198(26)$ \cite{Ade:2013zuv}.   Direct detection experiments place stringent bounds on the DM-Higgs coupling ${\mathcal L } \ni y_{\chi \chi h} (\bar{\chi} \chi h)$, $y_{\chi \chi h} \lsim 7 \times 10^{-3} (m_{X} / 50 \rm{\, GeV})^{1/2}$ for $m_\chi \gsim 50$ GeV \cite{Akerib:2016vxi,Basirnia:2016szw}.  These bounds make it difficult to realize the thermal abundance solely via a Higgs boson coupling, \ie, ``Higgs portal'' DM is constrained---see, \eg, \cite{Beniwal:2015sdl}.  It is therefore natural to consider the possibility where the thermal abundance is obtained absent a large coupling to the Higgs boson---perhaps solely via coupling to the $Z$, which induces only SD scattering.  However, as the xenon (Xe) nuclei of LUX and PandaX have spin, direct detection experiments also probe this scenario.\footnote{Additional bounds result from the lack of observation of neutrinos in the IceCube detector \cite{Aartsen:2016exj}, as produced via solar DM capture and subsequent annihilation, though these are generally weaker.}  As we will show, Majorana DM with thermal history primarily determined by $Z$ couplings is being probed now.

In this paper, we first discuss the simplest, gauge-invariant DM model wherein a Majorana fermion interacts with a $Z$ boson.  
We will see that this coupling generically induces a large contribution to the $\rho$ parameter. We then discuss how $Z$-mediated dark matter may be realized as a limit of the singlet-doublet model \cite{ArkaniHamed:2005yv}. This model has no problem with the $\rho$ parameter. However, while direct detection is primarily mediated via $Z$ boson exchange, other couplings may be important for the dark matter's thermal history.  We will demonstrate that a large region of parameter space in this model is close to being probed, even if the couplings to the Higgs boson that determine the SI interactions vanish.

\section{$Z$-mediated Dark Matter}

A gauge-invariant coupling of a Majorana fermion $\chi$ to the $Z$ may be generated via a higher-dimension operator involving the Higgs doublet $H$
\begin{equation}
\mathcal{L}\supset \frac{c}{2\Lambda^2}(i H^\dagger D_\mu H + {\rm h.c.})\bar{\chi}\gamma^\mu \gamma^5 \chi,
\label{eqn:Zop}
\end{equation}
where $c$ is a coupling constant and $\Lambda$ is the effective scale for new physics.  We have implemented this model in {\tt Micromegas v4.3.1} \cite{Belanger:2014vza}, which we use for calculations of relic density and direct detection processes.  This operator induces a coupling to the $Z$ boson (using $\vev{H} = \frac{v}{\sqrt{2}}$, $v= 246$ GeV)
\begin{equation}
\label{eqn:Zcoup}
\mathcal{L}\supset -\frac{g_2}{4 c_W}\frac{c v^2}{\Lambda^2}Z_\mu \bar{\chi}\gamma^\mu \gamma_5 \chi \equiv -\frac{g_2}{2 c_W} g_{A} Z_{\mu} \bar{\chi} \gamma^{\mu} \gamma_5 \chi,
\end{equation}
in addition to four- and five-point interactions between the DM and the $Z$ and $h$ bosons with related strength.  In terms of $m_\chi$ and $g_A$, one can calculate the relic density and direct detection rate.  The dominant direct detection signal is spin dependent through the nucleon effective operator $\bar{\chi} \gamma_\mu \gamma_5 \chi N \gamma^\mu \gamma_5 N$.  Other operators are velocity suppressed \cite{Fitzpatrick:2012ix,Anand:2013yka}.

Fermionic DM with purely axial vector coupling to the $Z$ was studied in \cite{Arcadi:2014lta}, whose results for the relic density we have reproduced, and more recently in \cite{Escudero:2016gzx}, whose results are in agreement with ours.\footnote{A discrepancy with the relic density calculation appearing in the original version of \cite{Escudero:2016gzx} has since been resolved in a later version.}
However, the model considered in these papers is not gauge invariant.  The gauge-invariant version was studied in \cite{deSimone:2014pda}, with closely related work in \cite{Belanger:2015hra}.  Our results appear consistent with \cite{Belanger:2015hra}, but our relic density calculation and direct detection limits differ from \cite{deSimone:2014pda}.

\begin{figure}[t]
\includegraphics[width=\columnwidth]{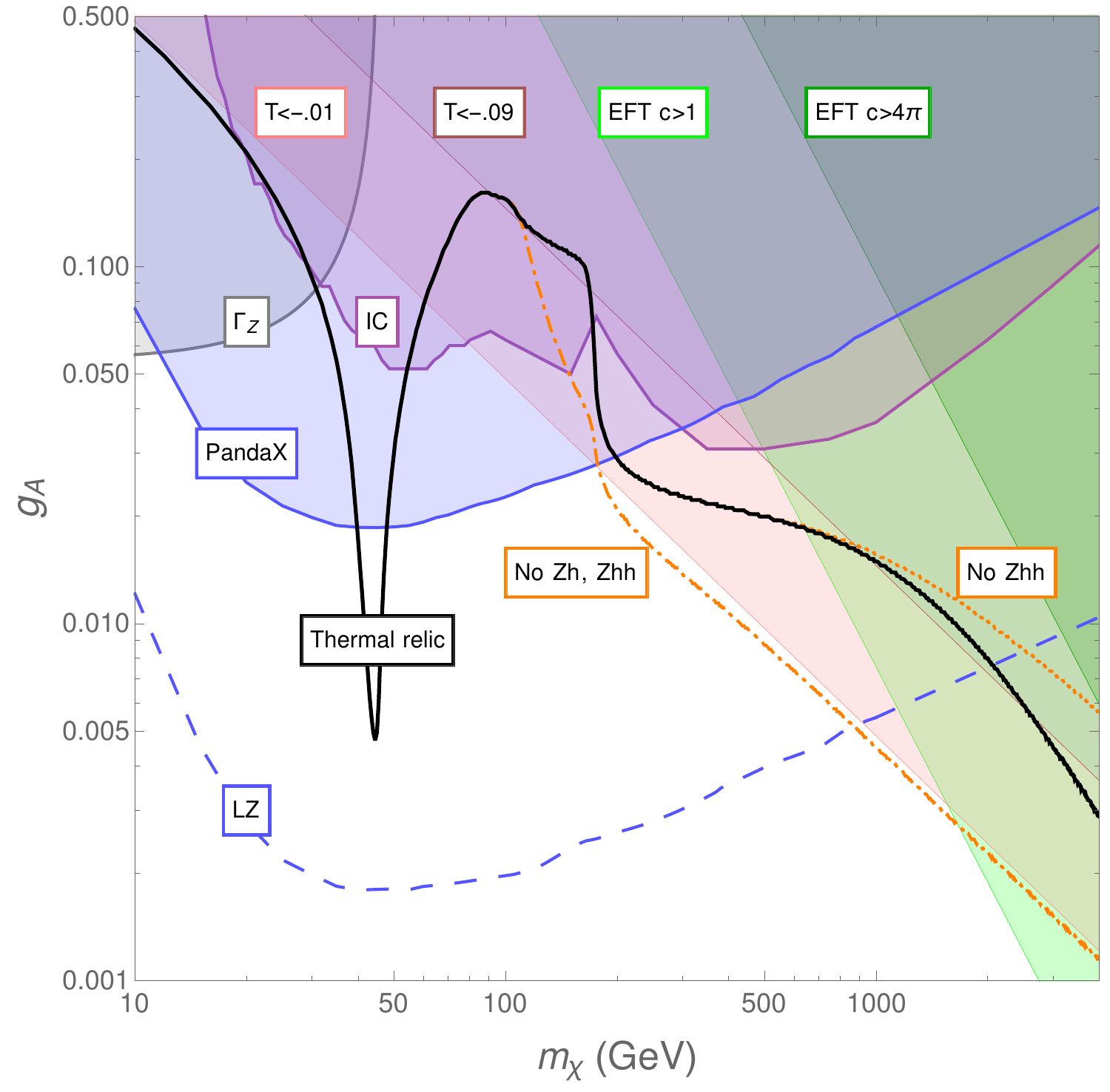}
\caption{Constraints on $g_A \equiv c v^2/(2 \Lambda^2)$ as a function of $m_\chi$ for $Z$-mediated DM.  The black solid line indicates where the thermal relic abundance matches the observed abundance \cite{Ade:2013zuv}.  The orange dotted (dot-dashed) line corresponds to omitting higher-dimensional couplings to $Zhh$ (and $Zh$), in violation of gauge invariance.  Shaded (dashed) blue is excluded by PandaX (LZ), with bounds obtained as described in the text.  The purple shaded region is excluded by IceCube (IC).  Invisible $Z$ decay limits ($\Gamma_Z$) are shown in gray. Pink shaded regions indicate where $T<-0.09$ (upper, darker) and $T<-0.01$ (lower, lighter), corresponding to $2\sigma$ excluded regions depending if $S$ is nonzero or zero, respectively. Green shaded regions indicate where  the EFT is precarious, taking $c_{\rm max}=4 \pi$ (upper, darker) or 1 (lower, lighter).}
\label{fig:Zportal}
\end{figure}

In \Fref{fig:Zportal}, we show (black line) the value of $g_{A}$ that reproduces the observed thermal relic density as a function of DM mass $m_\chi$.   Also shown (dot-dashed orange curve) is the required value if one were to introduce the coupling in (\ref{eqn:Zcoup}) without the attendant $\chi\chi Z h$ or $\chi \chi Z h h$ couplings, thereby violating gauge invariance (analytic results for this case are given in \cite{Arcadi:2014lta}).  Without the $\chi \chi Z h$ coupling, the annihilation to $Zh$ grows more rapidly as a function of $\sqrt{s}$.  Similarly, shown in dotted orange is the relic density calculation if only $2 \to 2$ annihilations are considered.  This neglects the $2 \to 3$ annihilation to $Zhh$, which becomes important at large $m_\chi$ where the cross section's mass dependence outweighs the phase space suppression.\footnote{To our knowledge, no other studies have considered these $2 \to 3$ annihilations.}

The blue shaded region is excluded by PandaX bounds on SD scattering \cite{Fu:2016ega}, assuming $\chi$ makes up all of the DM.
This excludes thermal relics with $m_\chi \lsim 200$ GeV from making up all of the DM aside from the resonance at $m_\chi=m_Z/2$.\footnote{Currently published SD bounds from LUX \cite{Akerib:2016lao} are slightly less sensitive than PandaX.  However, Ref.~\cite{Akerib:2016lao} does not include the full exposure of the most recent bound on SI scattering published by LUX \cite{Akerib:2016vxi}.  A rescaling of the bounds in \cite{Akerib:2016lao} including the longer exposure of \cite{Akerib:2016vxi} would exclude thermal relics with $m_\chi \lsim 240$ GeV, slightly stronger than the PandaX bound.}
Incidentally, the region above this curve with $m_\chi \lsim 200$ GeV is excluded even for thermally produced $\chi$ making up only a portion of dark matter because the bound scales as $\sSD \Omega \propto \sSD/\langle \sigma v \rangle \propto g_A^2/g_A^2$.  
Projected bounds from LZ \cite{Akerib:2015cja} are also shown (blue dashed).
Since projections for SD scattering bounds are not given by LZ, we estimate them by rescaling LUX results from \cite{Akerib:2016lao} by the same factor as SI bounds improve from \cite{Akerib:2015rjg}---which has the same exposure as \cite{Akerib:2016lao}---to the projected LZ SI bounds \cite{Akerib:2015cja}. 
LZ will probe thermal relics up to $m_\chi \lsim 2$ TeV.

IceCube bounds \cite{Aartsen:2016exj} on annihilations of DM captured in the sun by the $\chi$-proton SD cross section are shown in purple. To account for multiple annihilation final states, we estimate
\begin{equation}
\label{eqn:ICApprox}
\sigma_{\rm SD,p,bound}= \left(\sum\limits_{i={\rm channels}} \frac{{\rm Br}_i}{\sigma_i}\right)^{-1},
\end{equation}
where ${\rm Br}_i$ is the branching ratio to channel $i$ and $\sigma_i$ is the IceCube bound assuming 100\% branching ratio to channel $i$. At DM masses below the $W$ mass, annihilations to neutrinos and taus give the strongest bounds, while intermediate masses are most constrained by annihilations to tops and higher masses $\gsim {\rm TeV}$ by $Zhh$ annihilations.\footnote{We take limits from $Zhh$ to be roughly comparable to those from $ZZ$ and $hh$. While the neutrinos from a three-body final state will be less energetic, for large $m_\chi$ the energies are still expected to be above threshold. As such, the presence of additional neutrinos should lead to comparable (if not stronger) limits.} 
Although the approximation (\ref{eqn:ICApprox}) is not precise, we view it as suitable, particularly as sensitivities of Xe-based experiments and IceCube are currently only competitive at higher $m_\chi \gsim$ few hundred GeV, and Xe-based limits will soon dominate.  Constraints from searches for gamma-ray signals from DM annihilation \cite{Ahnen:2016qkx} are similarly subdominant.

For $m_\chi<m_Z/2$, $g_A$ is bounded by LEP measurements of the invisible $Z$ width, which limits $\Gamma(Z \to \chi \chi)<2$ MeV at 95\% confidence level \cite{ALEPH:2005ab}. This bound is shown in gray using
\begin{equation}
\Gamma(Z \to \chi \chi)=\frac{m_Z}{6 \pi} \left(g_A \frac{g_2}{2 c_W}\right)^2 \left(1-\frac{4 m_\chi^2}{m_Z^2}\right)^{3/2}.
\end{equation}
The LHC can probe larger DM masses than LEP with monojet-type searches; however, both present and future sensitivities will be subdominant to LEP or direct detection bounds \cite{deSimone:2014pda}, with the background $(Z \to \nu \nu)+{\rm jet}$ representing an important irreducible background.

While pure $Z$-mediated DM currently evades the above experimental constraints at larger masses, there are other considerations that should be taken into account when evaluating whether this is a reasonable benchmark model. First, the coupling of \eq{eqn:Zcoup} will generate a large contribution to the $\rho$ parameter.  At loop level, two insertions of the operator in \eq{eqn:Zop} generate a contribution to the self-energy of the $Z$ boson
\begin{equation}
\mathcal{\delta L}\supset \frac{c^2 m_\chi^2}{\pi^2 \Lambda^4} \log\left(\frac{\Lambda}{m_\chi}\right) |H^\dagger D_\mu H|^2,
\end{equation}
without a corresponding contribution to the $W$ boson self-energy.  This gives
\begin{equation}
\label{eqn:delrho}
\delta \rho= -\frac{c^2 m_\chi^2 v^2}{2 \pi^2 \Lambda^4}\log\left(\frac{\Lambda}{m_\chi}\right).
\end{equation}
Under the strong assumption that no other operators affecting electroweak precision physics are generated,  $T=\delta\rho/\alpha(M_Z)>-0.01$ at $2\sigma$ \cite{Agashe:2014kda}.  The corresponding constraint is shown in \Fref{fig:Zportal} as a light pink shaded region, taking the logarithm in \eq{eqn:delrho} to be unity.  If, however, nontrivial contributions to $S$ are simultaneously allowed (but $U=0$), $T>-0.09$ at $2\sigma$. This constraint is shown in dark pink.  It is clear there is tension between precision electroweak constraints and obtaining a thermal history in this model, particularly at high masses.\footnote{Writing down $|H^{\dagger} D_{\mu} H|^2$ directly, with $\Lambda$ suppression comparable to that in \eq{eqn:Zcoup}, would be even worse.  The estimate of \eq{eqn:delrho} corresponds to the idea there is an approximate custodial $SU(2)$ symmetry, broken only via the DM-$Z$ coupling.}

Another question is whether the model of \eq{eqn:Zcoup} is a valid effective field theory (EFT) for describing DM annihilations in the early univese.  The relevant scale for these annihilations is $2 m_{\chi}$.  
Without appreciable separation between $\Lambda$ and $m_{\chi}$ we expect higher-dimension operators to be important.  To illustrate regions where the EFT is precarious, we take
\begin{equation}
g_A \lsim 0.095 \left(\frac{c_{\rm max}}{4 \pi}\right) \left(\frac{\rm TeV}{m_\chi}\right)^2 \left(\frac{2 m_\chi}{\Lambda_{\rm min}}\right)^2, 
\end{equation}
where for illustrative purposes we have set $\Lambda_{\rm min}= 2 m_{\chi}$.  In the figure, we have shown two regions:  a light green one where we have set $c_{\rm max} =1$, and a dark green one where we have allowed $c_{\rm max} = 4 \pi$.  At large $m_\chi$, describing the physics with an EFT becomes more difficult.

There are two main takeaways from this section.
First, current direct detection bounds constrain thermal $Z$-mediated dark matter $\lsim 200$ GeV aside from a tiny window where annihilations are resonant.  For a recent discussion of possibilities of probing this region, see \cite{Hamaguchi:2015rxa}.  Increased sensitivity by next-generation experiments will probe higher masses near the limit of validity for the EFT.
Second, this coupling of the DM to the $Z$ boson maximally breaks custodial $SU(2)$.  As such, there is tension with constraints on the $\rho$ parameter for the entirety of the thermal relic space with $m_{\chi} > m_Z /2$, except perhaps at very large DM mass where the validity of the EFT is questionable.

\section{Embedding $Z$-mediated Dark Matter in the Singlet-Doublet Model}

A simple embedding of this $Z$-mediated DM model that moves beyond an EFT, drastically lessens the tension with the $\rho$ parameter, and is consistent with approximate gauge coupling unification is the singlet-doublet model \cite{ArkaniHamed:2005yv}.  Early analysis of this model appeared in \cite{Mahbubani:2005pt,D'Eramo:2007ga,Enberg:2007rp}, with focus on the direct detection phenomenology in \cite{Cohen:2011ec}. It is closely related to the DM story in split supersymmetry \cite{Pierce:2004mk,Giudice:2004tc} or the well-tempered neutralino \cite{ArkaniHamed:2006mb}.  More recent studies of the DM phenomenology appear in \cite{Cheung:2013dua,Banerjee:2016hsk,Basirnia:2016szw}.   Related collider studies appear in \cite{Calibbi:2015nha, Freitas:2015hsa}.

The singlet-doublet model is obtained by adding to the Standard Model a vectorlike pair of electroweak doublets $D$ and $D^c$ with hypercharge $Y = \pm \frac{1}{2}$ and an electroweak singlet $N$ with $Y = 0$.
The relevant interactions in the Lagrangian are
\be
{\cal L} \supset - y D H N - y^c D^c \tilde{H} N - M_D D D^c - \frac{M_N}{2} N^2 + \text{ h.c.}
\ee
The Yukawa couplings generate mixing between $N$ and the electromagnetically neutral components of $D, D^c$, giving rise to three Majorana fermions, the lightest of which is a DM candidate.
Because the DM is descended in part from an $SU(2)$ doublet, it will couple not only to the $Z$, but also to the $W$ boson. This generates a correction to the $W$ self-energy, which contributes to mitigating the constraints from $\rho$, but also generically affects the early universe cosmology.

The Majorana nature of the DM ensures that it does not exhibit vectorial couplings to the $Z$ boson, avoiding contributions to SI scattering that would be far in excess of current limits. 
Thus, the coupling to the $Z$ is of the same form as the right-hand side of \eq{eqn:Zcoup} but with the coupling determined by
\begin{equation}
g_A=\frac{1}{2} \frac{\Delta^2 v^2({y^{c}}^2-y^2)}{\Delta^2+v^2((y^2+{y^{c}}^2)(M_D^2+m_\chi^2)+4yy^c M_D m_\chi)},
\end{equation}
where $\Delta^2=M_D^2-m_\chi^2$ and $m_\chi$ is the DM mass determined by the mixing of $N$ and the neutral states in $D$ and $D^c$.

However, bounds on SI scattering are sufficiently strong that they also constrain DM that interacts via the Higgs boson. As such, it is of particular interest to consider this model in the so-called ``Higgs blind spot'' \cite{Cohen:2011ec,Cheung:2012qy,Cheung:2013dua}, wherein the coupling to the Higgs boson vanishes:
\be
\label{eq:bscondition}
y^c_{\rm BS} = - y \frac{M_N}{M_D} \left(1 \pm \sqrt{1 - \left(\frac{M_N}{M_D}\right)^2}\right)^{-1}.
\ee
In this blind spot, the DM will retain a diagonal coupling to the $Z$ boson (as in the previous section) but will also exhibit off-diagonal couplings to the $Z$ as well as to the $W$ boson. 
So, while the DM phenomenology in certain regions of the singlet-doublet parameter space will correspond to that of the $Z$-mediated case, these additional couplings can play a significant role elsewhere.

\begin{figure}[t]
\includegraphics[width=\columnwidth]{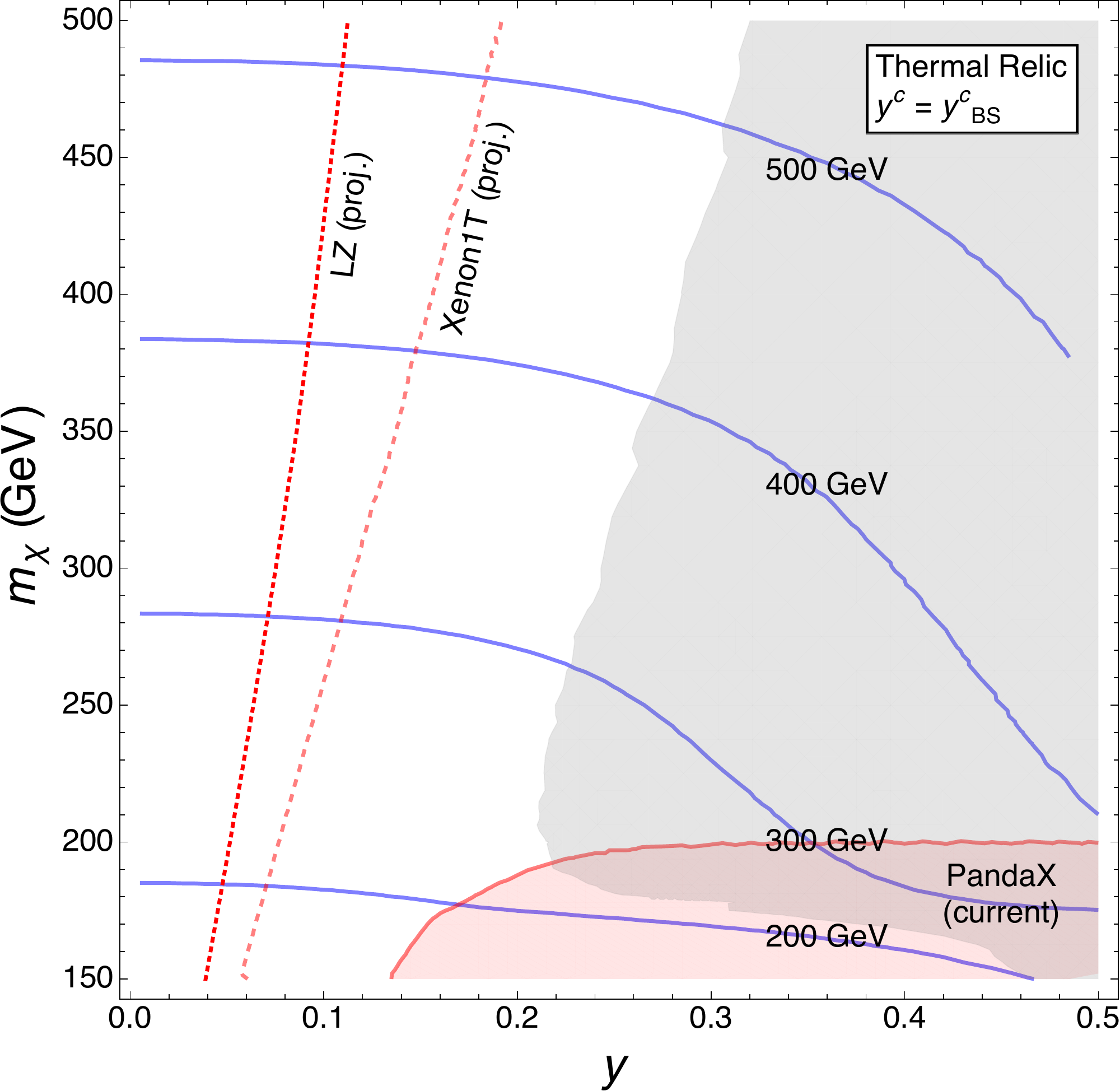}
\caption{Constraints on the singlet-doublet model at the Higgs blind spot.  The blind spot condition fixes $y^c$ via (\ref{eq:bscondition}), while $M_D$ (contours in blue) is set by demanding the thermal relic abundance matches the observed DM abundance.  Shown in red are current bounds from PandaX (solid curve, shaded region) and prospective bounds from Xenon1T (thick dashed curve) and LZ (thin dashed curve) on $\sSD$.  In the shaded gray region, the DM-$Z$ is at least 95\% of that in the $Z$-mediated model of the previous section.}
\label{fig:SDrelicfixed}
\end{figure}

In \Fref{fig:SDrelicfixed}, we have fixed two of the four free parameters of the model as follows: $y^c$ is fixed to the Higgs blind spot value, so that $\sSI$ vanishes at tree level, and $M_D$ is fixed to agree with the observed (thermal) relic density calculated using {\tt Micromegas}.\footnote{There is one physical phase among the parameters $\{y,y^c,M_D,M_N\}$, which for simplicity we set to zero. Effects of a nonzero phase are discussed in \cite{Mahbubani:2005pt,D'Eramo:2007ga}.}
We specialize to the regime $-\abs{y} < y^c < 0$, which corresponds to taking the plus solution in \eq{eq:bscondition}---the sign choice is simply a manifestation of the fact that the physics is left invariant by the exchange $y \leftrightarrow y^c$. For instance, if we were to write $\cos \theta = \frac{M_N}{M_D}$, \eq{eq:bscondition} could be rewritten as
\be
\frac{y^c_{\rm BS}}{y} = - \sqrt{\frac{1 \mp \sin \theta}{1 \pm \sin \theta}}.
\ee
Choosing the opposite sign would reproduce the plot with $y$ and $y^c$ exchanged.

Several quantities of interest are then plotted as a function of the remaining two free parameters, $y$ and the DM mass $m_\chi$ ($= M_N$ in the blind spot, see, \eg, \cite{Cohen:2011ec,Banerjee:2016hsk}).
The shaded red region is excluded by PandaX bounds on $\sSD$.  Also shown are projected bounds from Xenon1T \cite{Aprile:2015uzo} and LZ, which if no detection is made would exclude the regions to the right of each line.
Blue contours represent $M_D$, or equivalently the mass of the charged state.
We do not display bounds from IceCube because at present LUX and PandaX provide the strongest constraints throughout the parameter space shown, and as noted in the previous section direct detection experiments will scale in sensitivity much faster than IceCube.

The gray shaded region to the right of the plot indicates where the DM-$Z$ coupling is $\gsim 0.95$ of the value prescribed in the pure $Z$-mediated case. Here, the DM cosmology is well described by the simple $Z$-mediated model, but the additional states nearby in mass that fill out complete electroweak representations provide targets for collider searches and render contributions to the $T$ parameter small, as we now detail.

LHC searches for charginos and neutralinos decaying via electroweak bosons could probe this model.  However, current limits \cite{Aad:2014vma,Aad:2015jqa,CMS:2016gvu,CMS:2016zvj} are mild and do not appear in this region of parameter space.  Future limits likely will---see, \eg, \cite{ATL-PHYS-PUB-2015-032}.

Far into the gray region (at large $y$ beyond what is plotted), $\Delta T$ is similar in size to the expectation from the $Z$-mediated model above.
However, there are additional comparably sized contributions from the doublet, as the splitting within the doublet is related to the DM-$Z$ coupling.  Partial cancellation between these contributions leads to a value somewhat smaller than the na\"ive expectation. For instance, in the limit $M_D \gg m_\chi, yv$,
\be
\delta \rho = \frac{y^4 v^2}{48 \pi^2 M_D^2} \left\{1 + \frac{17 m_\chi^2}{4 M_D^2} - \frac{6 m_\chi^2}{M_D^2} \log\left(\frac{M_D}{m_\chi}\right)\right\}.
\ee
For larger $m_\chi/M_D$ (smaller $y$), higher-order terms are relevant and result in further suppression. As such, $\abs{\Delta T} \lsim 2 \times 10^{-3}$ throughout \Fref{fig:SDrelicfixed}.

To the left of the plot, away from where the $Z$-mediated description is sufficient, $M_D$ is not too much larger than $m_\chi$, such that coannihilation and $t$-channel annihilation to $W W$ become increasingly important in determining the relic density. Interestingly, future direct searches will be able to probe a sizable portion of this regime.
Though future direct detection will have trouble constraining the entire region, and the model will continue to evade these constraints for sufficiently small couplings, collider searches may provide a promising alternative probe. For instance, the small mass splittings make this region of parameter space susceptible to searches such as \cite{CMS:2016zvj} based on soft leptons.

So, while the pure $Z$-mediated model is a good proxy for the singlet-doublet model in the blind spot at large $y$ (where direct detection constraints may be directly translated), the reach of future experiments extends well beyond this regime to smaller $y$.
Even if the $Z$-mediated model of the preceding section is excluded up
to a given $m_\chi$, this model presents a minimal variation in which that DM mass
remains viable, and yet meaningful constraints can still be achieved.
As such, the singlet-doublet model in the Higgs blind spot represents a worthy target for future WIMP searches, and it would be valuable for experiments to quote constraints in terms of this parameter space.

\begin{figure}[t]
\includegraphics[width=\columnwidth]{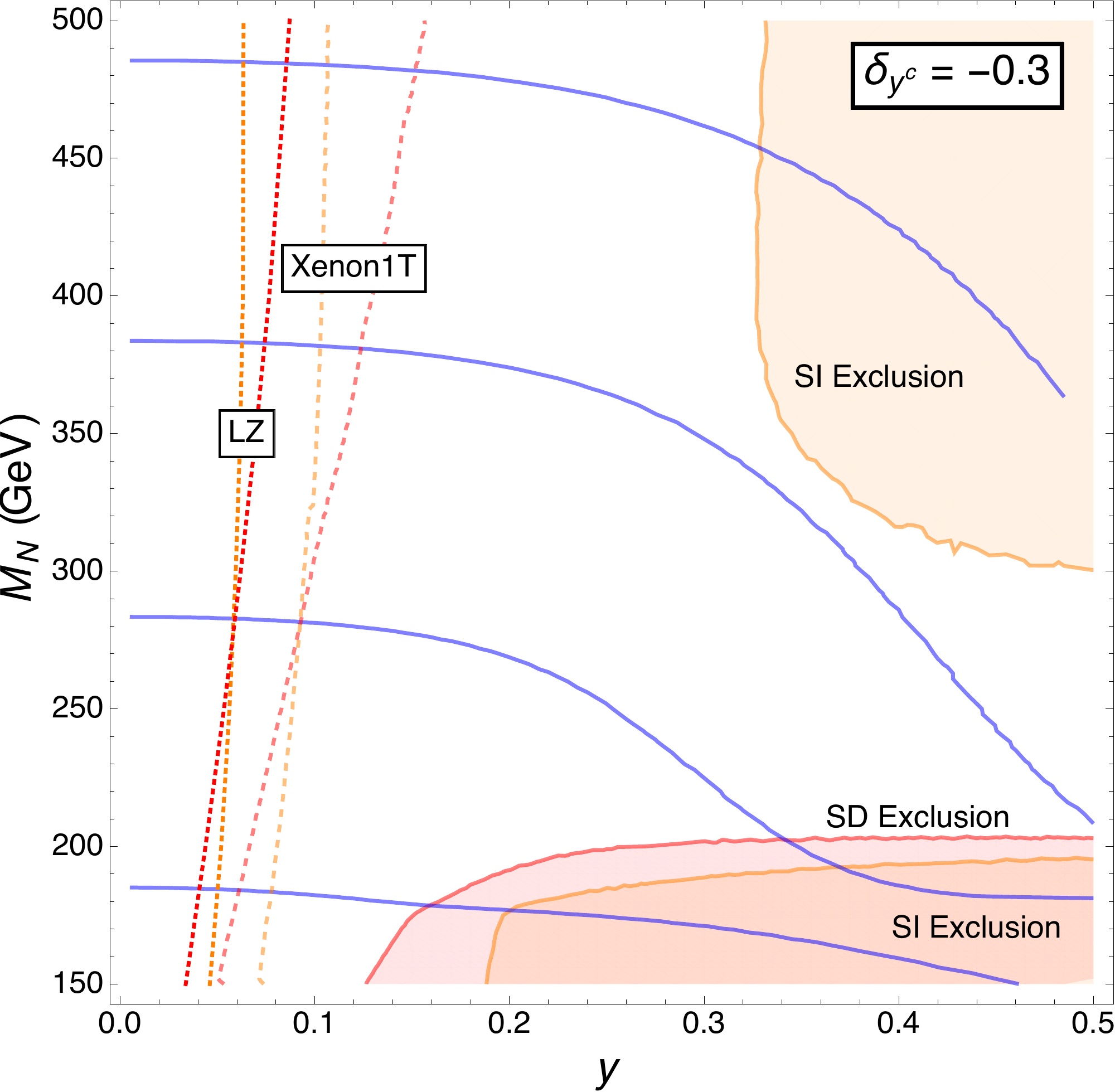}
\caption{Similar to \Fref{fig:SDrelicfixed}, but with $y^c$ deviating from the Higgs blind spot by $\delta_{y^c}=-0.3$, see \eq{eqn:deviation}.  Mirroring the previous figure, contours of $M_D$ are shown in blue. Current and projected limits on $\sSD$ ($\sSI$) are shown in red (orange).}
\label{fig:nonBS}
\end{figure}

Finally, we comment on the effect of tuning away from the exact blind spot. This will result in a nonzero DM-$h$ coupling that, as mentioned previously, can result in strong constraints from SI scattering---in fact, this is the case even if the DM-$h$ coupling is sufficiently small to have a negligible impact on the DM thermal history. Parameterizing the deviation from the blind spot as
\be
\label{eqn:deviation}
\delta_{y^c} = \frac{y^c}{y^c_{\rm BS}} - 1,
\ee
we show in \Fref{fig:nonBS} the parameter space for $\delta_{y^c} = -0.3$. Note that, for $y^c \neq y^c_{\rm BS}$, $m_\chi \neq M_N$, so here we plot with respect to $\{y, M_N\}$---however, for the values of $y$ and $\delta_{y^c}$ considered, $m_\chi \simeq M_N$.

To the left, where $t$-channel and co-annihilation play a significant role in determining the relic density, changing $y^c$ simply changes the DM-$Z$ and DM-$h$ couplings, altering the exact values of $\sigma_{\rm SD, SI}$ (and hence the future experimental reach) relative to \Fref{fig:SDrelicfixed} in this region of parameter space.
To the right however, where the relic density is predominantly determined by $s$-channel $Z$ exchange, $M_D$ changes to compensate the change in $y^c$ and maintain approximately the same DM-$Z$ coupling as in \Fref{fig:SDrelicfixed}. As a result, the current SD exclusions (and the region well described by the pure $Z$-mediated model) do not change significantly.
While it is not plotted, we note that the parameter space is not symmetric about $y^c_{\rm BS}$.

The two disjoint regions at larger $y \gsim 0.35$ are related to the top quark threshold. For $m_\chi \gsim m_t$, the observed relic abundance is achieved for a smaller DM-$Z$ coupling, which is accompanied by a similar suppression of the DM-$h$ coupling. This allows the model to evade present SI limits near threshold ($m_{\chi} \gsim$ 190 GeV). 
However, while the relic density constraint largely fixes the DM-$Z$ coupling at larger $M_N$, the DM-$h$ coupling exhibits different parametric dependence and modestly increases with $M_N$.  Eventually this increase results in a second excluded region for $m_{\chi} \gsim 300$ GeV.
For $\delta_{y^c} = -0.3$, $\sSI$ for $200 \GeV \lsim M_N \lsim 300 \GeV$ is just below current LUX limits while, for larger $\delta_{y^c}$, this gap does not appear.

For the value of $\delta_{y^c}$ shown, constraints from SI and SD scattering are complementary today, excluding slightly different regions of parameter space, and are comparable in the future.  This point was chosen specifically to show where the future constraints may be roughly similar.\footnote{Because the SI and SD constraints will be comparable in the future for the $\delta_{y^c}$ in \Fref{fig:nonBS}, the exclusions should be combined to yield a somewhat stronger bound on $y$.  We have chosen not to do so in order to demonstrate the relative strength of each, and anyway for most values of $\delta_{y^c}$ only one of the SI or SD constraints will dominate.}
For larger $\delta_{y^c}$, SI constraints rapidly dominate, \eg,  already excluding much of the parameter space for $\abs{\delta_{y^c}} = 0.5$ while still not significantly altering the thermal history. For $\abs{\delta_{y^c}} \lsim 0.1$, SD constraints dominate throughout the parameter space.
But, in all cases, an order of magnitude improvement in limits would require the model to lie squarely in the small coupling and coannihilation regime, though the exact regions of parameter space probed will depend on the proximity of $y^c$ to the blind spot value.

Incidentally, there are two notable regions of singlet-doublet parameter space that provide a thermal DM candidate via coannihilation but with suppressed direct detection cross sections.
First, in the Higgs blind spot discussed here, there is a ``double blind spot'' where both of the DM-$h$ and DM-$Z$ couplings vanish. This occurs at $M_N \simeq M_D \simeq 880$ GeV (with a slight $y \simeq y^c$ dependence).
In this case, there is a single doublet-like Majorana particle degenerate with the chargino.
The second interesting case is that of the nearly pure doublet ``Higgsino'' when $M_N \gg M_D$.  The neutral components of the doublets are split into a pseudo-Dirac state, suppressing direct detection cross sections, and the observed relic abundance is obtained for $M_D \simeq 1.1$ TeV.

\section{Conclusions}

A simplified DM EFT in which the dark matter communicates with the SM through a $Z$ boson represents a valuable target for WIMP searches. However, recent improvements in direct detection limits have begun to force the simplest gauge-invariant version of such a model into a region of parameter space exhibiting large contributions to precision electroweak parameters and in which one might question the validity of the EFT. As such, the time is ripe to consider how models of WIMPs beyond the simplest examples fare in the face of current and future direct searches.

In this paper, we have discussed the singlet-doublet model, which exhibits similar phenomenology to the $Z$-mediated model in certain regions of parameter space. However, the presence of additional states nearby in mass that fill out complete electroweak representations prevents overly large contributions to precision electroweak parameters. Moreover, the extra DM couplings, notably to the $W$ boson, allowed by these additional states lead to different DM phenomenology. In particular, contributions to the DM annihilation cross section from $t$-channel DM partner exchange or coannihilation can allow the correct thermal relic density to be achieved with a small DM-$Z$ coupling, opening new regions of parameter space that represent exciting targets for future experiments.  Moreover, the additional partners of the DM could be discovered at the high-luminosity LHC.

Future direct detection will probe well beyond where the DM cosmology is described by the simplified $Z$-mediated model, and null results would allow only the case in which the Yukawa couplings are relatively small and the thermal relic density is achieved through coannihilation.
This would require a somewhat striking coincidence of parameters (which could perhaps arise from renormalization group fixed ratios as in \cite{Kearney:2013xwa}), with not only the mass of the charged state lying close to the DM mass but also the Yukawa couplings conspiring such that DM-$h$ coupling approximately vanished to evade the very stringent SI scattering constraints.
As limits continue to improve, alternatives to the simplest WIMP models, or even to WIMPs themselves, will become increasingly attractive objects for study.

\vspace{-0.2cm}
\subsection*{Acknowledgments}
\vspace{-0.3cm}
We thank Josh Ruderman and Mariangela Lisanti for useful conversations.
The work of NO and AP is supported by the U.S.~Department of Energy under Grant No.~DE-SC0007859.
JK is supported by the DOE under Contract No.~DE-SC0007859 and Fermilab, operated by Fermi Research Alliance, LLC under Contract No.~DE-AC02-07CH11359 with the United States Department of Energy.
This work was performed in part at the Aspen Center for Physics, which is supported by National Science Foundation Grant No.~PHY-1066293.

\bibliography{SDRefs}
\end{document}